\documentclass[aps,prd,amsmath,amssymb,twocolumn,preprintnumbers,nofootinbib]{revtex4-2}

\usepackage[T1]{fontenc}
\usepackage{times}
\DeclareMathAlphabet{\MATHIT}{OT1}{ptm}{m}{it}%% similiar to mathptmx
\DeclareSymbolFont{Letters}{OML}{ztmcm}{m}{it}%% dito
\DeclareSymbolFontAlphabet{\mathNormal}{Letters}% dito
\usepackage[utf8]{inputenc}
\usepackage[english]{babel}
\usepackage{amsmath}
\usepackage{amsfonts}
\usepackage{amssymb}
\usepackage{amsthm}
\usepackage{enumitem}
\usepackage{dcolumn}
\usepackage{bm}
\usepackage{bbm}
\usepackage{cancel}
\usepackage{tikz}
\usetikzlibrary{calc,shapes,arrows,patterns,decorations.pathmorphing,positioning,automata}
\usepackage{tabularx}
\usepackage{booktabs}
\usepackage{tensor}
\usepackage{scalerel}
\usepackage{xcolor}
\usepackage{mathtools}
\usepackage{float}
\usepackage{placeins}
\usepackage{xfrac}
\usepackage[pdfborder={0 0 0}]{hyperref}
\definecolor{darkblue}{rgb}{0,0,.5}
\definecolor{darkgreen}{rgb}{0,0.5,.5}
\definecolor{darkyellow}{rgb}{0.5,0.5,0}
\definecolor{fhl}{rgb}{1,0,0}
\hypersetup{colorlinks=true, breaklinks=true, linkcolor=darkblue, menucolor=darkblue, urlcolor=darkblue, linktocpage=true}
\usepackage{geometry}
\usepackage{textpos}
\usepackage{relsize}
\usepackage[titletoc,title]{appendix}
\allowdisplaybreaks
\usepackage{slashed}
\usepackage{natbib}
\usepackage[export]{adjustbox}

\makeatletter 
\newsavebox\myboxA 
\newsavebox\myboxB 
\newlength\mylenA 
\newcommand*\xoverline[2][0.75]{% 
    \sbox{\myboxA}{$\m@th#2$}%
    \setbox\myboxB\null% Phantom box 
    \ht\myboxB=\ht\myboxA% 
    \dp\myboxB=\dp\myboxA% 
    \wd\myboxB=#1\wd\myboxA% Scale phantom 
    \sbox\myboxB{$\m@th\overline{\copy\myboxB}$}%  Overlined phantom 
    \setlength\mylenA{\the\wd\myboxA}%   calc width diff 
    \addtolength\mylenA{-\the\wd\myboxB}% 
    \ifdim\wd\myboxB<\wd\myboxA% 
       \rlap{\hskip 0.5\mylenA\usebox\myboxB}{\usebox\myboxA}% 
    \else 
        \hskip -0.5\mylenA\rlap{\usebox\myboxA}{\hskip 0.5\mylenA\usebox\myboxB}% 
    \fi}
\makeatother

\let\originalleft\left
\let\originalright\right
\renewcommand{\left}{\mathopen{}\mathclose\bgroup\originalleft}
\renewcommand{\right}{\aftergroup\egroup\originalright}
\newcommand{\SU}[1]{\operatorname{SU}\left(#1\right)}
\newcommand{\su}[1]{\mathfrak{su}\left(#1\right)}
\newcommand{\of}[1]{\left(#1\right)}

\newcommand{\sof}[1]{\bigl(\big.#1\big.\bigr)}
\newcommand{\ssof}[1]{(#1)}
\newcommand{\fof}[1]{\left[#1\right]}

\newcommand{\cof}[1]{\left\{#1\right\}}

\newcommand{\avof}[1]{\left\langle #1\right\rangle}

\renewcommand*\[{\begin{equation}}
\renewcommand*\]{\end{equation}}
\newcommand{\order}{\mathcal{O}}

\newcommand{\ii}{\mathrm{i}}
\renewcommand*\Re{\operatorname{Re}}

\newcommand{\Tr}{\operatorname{Tr}}
\newcommand{\id}{\mathbbm{1}}

\renewcommand*\bar[1]{\ThisStyle{\xoverline{\SavedStyle #1}}}

\renewcommand*\hat[1]{\widehat{#1}}

\definecolor{newgreen}{RGB}{10,100,20}

\newcommand{\kr}[1]{\textcolor{newgreen}{#1}}

\definecolor{chcol}{RGB}{200,0,0}

\begin{document}

%%%%%%%%%%%%%%%%%%%%%%%%%%%%%%%%%%%%%%%%%%%%%%%%%%%%%%%%%%%%%%%%%%%%%%%%%%%

\title{Improved Dirichlet boundary conditions for lattice gauge-fermion theories}

\author{Tobias Rindlisbacher}
\email{trindlis@itp.unibe.ch}
\affiliation{Albert Einstein Center for Fundamental Physics \& Institute for Theoretical Physics, University of Bern, Sidlerstrasse 5, CH-3012 Bern, Switzerland}
\author{Kari Rummukainen}
\email{kari.rummukainen@helsinki.fi}
\affiliation{Department of Physics \& Helsinki Institute of Physics, University of Helsinki, P.O. Box 64, FI-00014 University of Helsinki, Finland}
\author{Ahmed Salami}
\email{ahmed.salami@helsinki.fi}
\affiliation{Department of Physics \& Helsinki Institute of Physics, University of Helsinki, P.O. Box 64, FI-00014 University of Helsinki, Finland}

\begin{abstract}
Hybrid Monte Carlo (HMC) simulations of lattice gauge theories with fermionic matter rely on the invertibility of the lattice Dirac operator. Near-zero modes of the latter can therefore significantly slow down the update algorithm and cause instabilities. This is in particular a problem when dealing with massless fermions. Homogeneous temporal Dirichlet boundary conditions can be used to remove zero modes from massless lattice Dirac operators, but the standard implementation of these boundary conditions can cause severe finite-volume cutoff effects in regions of parameter space where the physics at the ultraviolet (UV) cutoff scale is dominated by the fermionic instead of the gauge action. In lattice quantum chromodynamics (QCD) this is usually not an issue, as the gauge action dominates the UV physics and the problem does not show up. In studies of beyond standard model (BSM) theories, on the other hand, the finite-volume artifacts can be severe. We have identified the origin of these IR cutoff effects and propose a simple improvement on the homogeneous temporal Dirichlet boundary conditions to prevent them. We demonstrate the benefits of using our improved boundary conditions at the example of $\SU{2}$ lattice gauge theory with $N_f=24$ massless Wilson-clover flavors. Due to the large number of fermions in this theory, the boundary-related finite volume artifacts are particularly strong, and the effect from switching from the normal to our improved homogeneous Dirichlet boundary conditions is therefore distinct. 

\end{abstract}
\maketitle

\section{Introduction}\label{sec:intro}
Temporal Dirichlet boundary conditions (TDBs) are widely used in lattice studies of quantum chromodynamics (QCD) and related quantum field theories (QFTs). The corresponding Euclidean path integrals are known as "Schr\"odinger functionals" (SF) and have certain appealing features: appropriately chosen boundary conditions induce a background field in the bulk, which can be used to define a non-perturbative running coupling~\cite{Luscher:1992an}, the so-called SF coupling. Furthermore, the use of TDBs removes zero modes from massless lattice Dirac operator~\cite{Sint:1993un}, which is important for hybrid Monte Carlo (HMC) simulations of gauge-fermion theories, as the HMC update method relies on the invertibility of the lattice Dirac operator, and becomes unstable if the latter has near-zero eigenvalues.

TDBs for pure $\SU{N}$ lattice gauge theories were introduced in ref.~\cite{Luscher:1992an} and extended to incorporate Wilson-Dirac~\cite{Wilson:1974sk} and $\order\ssof{a}$ improved Wilson-clover~\cite{Sheikholeslami:1985ij} fermions in ref.~\cite{Sint:1993un}. Here $a$ is the lattice spacing. Straightforward implementation of TDBs gives rise to $\order\ssof{a}$ cutoff effects, which can be compensated by introducing boundary counterterms for the fermion and gauge part of the action. The coefficients of the counterterms have been evaluated in perturbation theory for SU(2) and SU(3) pure gauge theory \cite{Luscher:1992an,Luscher:1993gh} and with fundamental representation fermions \cite{Sint:1995ch,Luscher:1996vw,Luscher:1996sc}.
These calculations were extended to higher fermion representations in refs. \cite{Karavirta:2011mv,Karavirta:2012qd} and to higher SU($N$) groups in ref.~\cite{Hietanen:2014lha}.  The boundary improvement for 
Symanzik improved gauge action and clover fermions
%~\cite{Weisz:1982zw,Iwasaki:1983iya,Luscher:1984xn,Sheikholeslami:1985ij} 
was discussed in~\cite{Klassen:1997jf,Aoki:1998qd,Takeda:2003he}.

Since the introduction of gradient flow (GF) techniques~\cite{Luscher:2010iy} in lattice studies, the GF coupling proposed in~\cite{Fritzsch:2013je} has to a large extent replaced the background field SF coupling as the preferred definition of a non-perturbative running coupling on the lattice. SF boundary conditions are, however, still widely used in lattice studies of gauge theories with massless fermions in order to prevent instabilities due to near-zero Dirac-eigenvalues. 

The improvement of the GF in combination with SF boundary conditions is discussed in~\cite{DallaBrida:2016kgh}. Unfortunately these improvement techniques rely on perturbation theory and are applicable only if the bare lattice gauge coupling is sufficiently small. In QCD-like theories the gauge coupling becomes small as the distance scale is reduced, and hence the bare lattice gauge coupling can be kept small. However, large bare gauge couplings are met in theories where the coupling runs slowly, for example in studies of theories featuring an infrared fixed point.

In the non-perturbative regime, the naive use of unimproved SF boundary conditions can cause severe lattice artifacts, in particular if the physics at the UV cutoff scale is not dominated by the gauge action but significantly affected by the presence of fermion fields. While these artifacts, as we will illustrate, seem to be caused by the UV-details of the lattice implementation of the TDBs, they manifest themselves most prominently in the form of severe infrared-cutoff effects in the gradient flow (see Fig.~\ref{fig:gffinitevolartifacts}), and can therefore have implications for studies of the non-perturbative running of the coupling in gauge-fermion theories.  In this paper we present a straightforward improvement of standard TDBs, and show that these remove the harmful lattice artifacts.

The remainder of this paper is organized as follows: Sec.~\ref{sec:latticeform}  introduces the lattice actions and observables that will be used in the forthcoming sections. In Sec.~\ref{sec:dirichletbcsc} we discuss the standard implementation of homogeneous TDBs for $\SU{N}$ lattice gauge theories with Wilson-Dirac fermions. We explain why this implementation can give rise to severe infrared (IR) cutoff effects and how these manifest in gradient flow studies. In Sec.~\ref{sec:imprdirichletbc} we introduce the improved boundary conditions and show that the discussed cutoff effects are removed. Concluding remarks and a brief outlook to future studies are given in Sec.~\ref{sec:conclusion}.

\section{Lattice formalism}\label{sec:latticeform}
In this section we provide some details on the utilized lattice actions and observables. All quantities are dimensionless, meaning that the lattice representation of physical quantities with a non-trivial mass dimension are implicitly multiplied with an appropriate power of the lattice spacing, $a$, to render the product dimensionless. For a mass or energy, we have for example: $m=a\,\tilde{m}$ and $E=a\,\tilde{E}$; for the field strength: $F_{\mu\nu}=a^2\,\tilde{F}_{\mu\nu}$, and for a position vector: $x_{\mu}=\tilde{x}_{\mu}/a$, and so on, where the quantities with a tilde are the non-lattice ones, which have their natural mass dimension.

\subsection{Lattice actions}\label{ssec:latticeactions}
In this section we describe the lattice gauge and fermion actions used to perform the HMC updates. For the gauge part we use Wilson's $\SU{N}$ lattice gauge action~\cite{Wilson:1974sk},
\[
S_{G}=\frac{\beta_L}{N}\,\sum\limits_{x}\sum\limits_{\mu<\nu}\Re\Tr\of{\id-U_{\mu\nu}\of{x}}\ ,\label{eq:wilsongaugeaction}
\]
with $\beta_L=2\,N/g_0^2$ being the bare inverse gauge coupling and $U_{\mu\nu}\of{x}\in\SU{N}$ are the so-called plaquette variables, given by
\[
U_{\mu\nu}\of{x}=U_{\mu}\of{x}U_{\nu}\of{x+\hat{\mu}}U^{\dagger}_{\mu}\of{x+\hat{\nu}}U^{\dagger}_{\nu}\of{x}\ .\label{eq:plaquettevar}
\]
The link variables, $U_{\mu}\of{x}\in\SU{N}$, can be written as
\[
U_{\mu}\of{x}=\exp\sof{\ii\,A_{\mu}\of{x}}\ ,
\]
with $A_{\mu}\of{x}\in\su{N}$ being the $\mu$-component of the lattice gauge field, which is assumed to be constant along the link that connects the site $x$ to its nearest neighboring site in $\mu$-direction, $x+\hat{\mu}$. 

%\subsection{Lattice fermion action}\label{ssec:latticefermionaction}
For the fermion part we use the Wilson-clover action~\cite{Sheikholeslami:1985ij},
\[
S_F\,=\,\sum_{f=1}^{N_f}\,\sum_{x,y}\,\bar{\psi}^{\of{f}}_{x}\of{D^{\of{f}}_{x,y}\fof{U}+C^{\of{f}}_{x,y}\fof{U}}\psi^{\of{f}}_{y}\ ,\label{eq:wilsonfermionaction}
\]
where for each flavor $f$, $D^{\of{f}}\fof{U}$ is the Wilson-Dirac operator for the given set of gauge link variables, $U=\cof{U_{\mu}\of{x}}_{\forall x,\mu}$, and $C^{\of{f}}\fof{U}$ is the corresponding clover operator. The two operator can be written as,
\begin{multline}
D^{\of{f}}_{x,y}\fof{U}=\delta_{x,y}\id-\kappa_{f}\sum\limits_{\mu=1}^{4}\sof{\delta_{x+\hat{\mu},y}\of{\id-\gamma_{\mu}}U_{\mu}\ssof{x}\\
+\delta_{x-\hat{\mu},y}\of{\id+\gamma_{\mu}}U^{\dagger}_{\nu}\ssof{x-\hat{\mu}}}\label{eq:wilsondiracop}\ ,
\end{multline}
and
\[
C^{\of{f}}_{x,y}\fof{U}=c_{SW}\,\kappa_{f}\,\delta_{x,y}\frac{1}{2}\sum_{\mu<\nu}\fof{\gamma_{\mu},\gamma_{\nu}}\,P_{\mu \nu}\of{x}\ .\label{eq:cloverop}
\]
In Eqs.~\eqref{eq:wilsondiracop} and \eqref{eq:cloverop} $\kappa_{f}=1/(2\,m_{f,0}+8)$ is the hopping parameter for the flavor $f$, encoding the flavor's bare fermion mass, $m_{f,0}$, the parameter $c_{SW}=1+\order\of{g^2_0}$ is the Sheikholeslami-Wohlert coefficient, and $P_{\mu\nu}\of{x}$ the clover term~\cite{Sheikholeslami:1985ij}, defined as,
\begin{multline}
P_{\mu\nu}\of{x}=\frac{1}{4}\sof{U_{\mu}\of{x} U_{\nu}\of{x+\hat{\mu}} U^{\dagger}_{\mu}\of{x+\hat{\nu}} U^{\dagger}_{\nu}\of{x}\\
-U^{\dagger}_{\nu}\of{x-\hat{\nu}} U^{\dagger}_{\mu}\of{x-\hat{\mu}-\hat{\nu}} U_{\nu}\of{x-\hat{\mu}-\hat{\nu}} U_{\mu}\of{x-\hat{\mu}}\\
+U_{\nu}\of{x} U^{\dagger}_{\mu}\of{x-\hat{\mu}+\hat{\nu}} U^{\dagger}_{\nu}\of{x-\hat{\mu}} U_{\mu}\of{x-\hat{\mu}}\\
-U_{\mu}\of{x} U^{\dagger}_{\nu}\of{x+\hat{\mu}-\hat{\nu}} U^{\dagger}_{\mu}\of{x-\hat{\nu}} U_{\nu}\of{x-\hat{\nu}}}\ ,\label{eq:cloverterm}
\end{multline}
and which is a lattice approximation of the field strength tensor $F_{\mu\nu}\of{x}=P_{\mu\nu}\of{x}+\order\ssof{a^2}$.

\subsection{Gradient flow and observables}\label{ssec:observables}

The gradient flow~\cite{Luscher:2010iy} of the gauge field is performed with the L\"uscher-Weisz action~\cite{Luscher:1984xn} and we monitor three different observables of the flow-evolved gauge field as functions of the flow time $t$ resp. the corresponding flow scale $\lambda=\sqrt{8\,t}$. 
The first of these observables is the rescaled plaquette action density,
\[
\avof{s_{U}\of{x,t}}=t^2\,\avof{\Re\Tr\of{\id-U_{\mu\nu}\of{x,t}}}/N\ ,\label{eq:plqsdens}
\]
where $U_{\mu\nu}\of{x,t}$ is the plaquette variable from Eq.~\eqref{eq:plaquettevar} for the flow-evolved gauge link variables at flow time $t$. The second observable is the topological charge or instanton density,
\[
\avof{q_{\mathrm{topo}}\of{x,t}}=\frac{1}{32\,\pi^2} \epsilon_{\mu\nu\rho\sigma} \avof{\Tr\of{P_{\mu\nu}\of{x,t} P_{\rho\sigma}\of{x,t}}}\ ,\label{eq:topodens}
\]
with $P_{\rho\sigma}\of{x,t}$ being the clover term from Eq.~\eqref{eq:cloverop} for the flow-evolved gauge link variables at flow time $t$. The third observable is the gradient flow gauge coupling~\cite{Luscher:2010iy}, defined as
\begin{equation}
    g^2_{\mathrm{GF}}\of{\lambda} = \frac{2 \pi^2 \lambda^4 \avof{E\of{\lambda}}}{3\of{N^2-1}\of{1 + \delta_{L/a}\of{\lambda/L}}},
    \label{eq:gradientflow}
\end{equation}
with $\avof{E\of{\lambda}}$ being the average clover gauge action density at flow time $t$ and $\delta_{L/a}(\lambda/L)$ a finite-volume correction factor \cite{Fodor:2012td,Fritzsch:2013je}. To minimize boundary effects from the TDBs, the action density $\avof{E\of{\lambda}}$ is evaluated on the hyperplane where $x_4=N_t/2$.
We will also show results for the topological susceptibility,
\[
\chi_{\mathrm{topo}}=\avof{Q_{\mathrm{topo}}^2}-\avof{Q_{\mathrm{topo}}}^2\ , \label{eq:toposusc}
\]
with the topological charge given in terms of Eq.~\eqref{eq:topodens} as,
\[
Q_{\mathrm{topo}}=\sum\limits_{x}\,q_{\mathrm{topo}}\of{x,t_{L/2}}\ ,
\]
where $t_{L/2}$ is the flow time $t$ for which the flow scale $\lambda=\sqrt{8\,t}$ equals $L/2$.

\section{Homogeneous temporal Dirichlet boundary conditions}\label{sec:dirichletbcsc}

\begin{figure}[h]
\centering
\begin{tikzpicture}[scale=0.4,nodes={inner sep=0}]
  \pgfpointtransformed{\pgfpointxy{1}{1}};
  \pgfgetlastxy{\vx}{\vy}
  \begin{scope}[node distance=\vx and \vy]
    \foreach \i in {0,...,7} {
        \draw [thin,gray] (\i,0) -- (\i,8)  node[solid,black,above] at (\i,7.3) {};
    }
    \foreach \i in {0,...,8} {
        \draw [thin,gray] (0,\i) -- (8,\i) node[solid,black,left] at (0.7,\i) {};
    }
        
    \draw[very thick,black] (0,0) -- (8,0) node[black,right=0.1,pos=1,scale=1.1] {$x_4=0$};
    \draw[very thick,black] (0,8) -- (8,8) node[black,right=0.1,pos=1,scale=1.1] {$x_4=N_t$};

    \foreach \i in {0,...,7} {
        \draw [very thick,blue!80!black] (\i+0.05,8) -- (\i+0.95,8);
    }
    \foreach \i in {0,...,7} {
        \node[circle,fill=red!80!black,minimum size=5pt] at (\i,8) {};
    }    
    \foreach \i in {0,...,7} {
        \draw [very thick,blue!80!black] (\i+0.05,0) -- (\i+0.95,0);
    }
    \foreach \i in {0,...,7} {
        \node[circle,fill=red!80!black,minimum size=5pt] at (\i,0) {};
    }    
    
    \draw[->] (-1,0) -- (-1,2) node[black,thick,left=0.1,pos=0.5,scale=1.1] {$x_4$};
    \draw[->] (0,-1) -- (2,-1) node[black,thick,below=0.1,pos=0.5,scale=1.1] {$x_i$};
  \end{scope}
\end{tikzpicture}
\caption{Illustration of (ordinary) homogeneous temporal Dirichlet boundary conditions (TDBs) with trivial boundary fields: spatial gauge links within the boundary layers at $x_4=0,N_{t}$ (blue links) are set to the identity, while the fermion fields are required to vanish on the boundary (red dots).}
\label{fig:normalhomdiribc}
\end{figure}
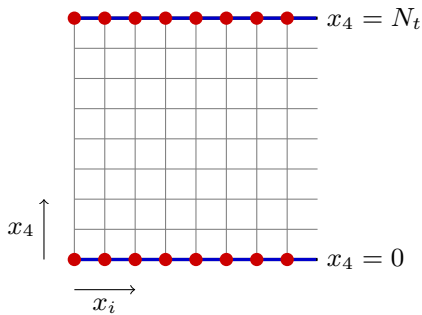

A typical implementation of homogeneous TDBs in $\SU{N}$ lattice gauge theory with Wilson-Dirac fermions, described by the Dirac operator in Eq.~\eqref{eq:wilsondiracop}, is illustrated in Fig.~\ref{fig:normalhomdiribc} for the case of trivial boundary fields: the spatial gauge links within the boundary layers at $x_4=0$ and $x_4=N_{t}$, where $N_t$ is the temporal size of the lattice, are set to the identity, and the fermion fields are required to vanish on the boundary\footnote{The implementation of TDBs for Wilson-Dirac fermions, as presented in~\cite{Sint:1993un}, requires in principle only to fix half of the fermionic degrees of freedom on the boundary. However, as also the gauge field is held fixed along the boundary, the remaining fermionic boundary degrees of freedom have no opportunity to interact with the bulk system and can therefore, without loss of generality, be set to zero.}. This is sufficient to remove near-zero modes from the Wilson-Dirac operator~\cite{Sint:1993un} and to thereby stabilize the HMC update algorithm.

For QCD or QCD-like lattice theories, where the physics at the UV-cutoff scale is for most of the simulation parameter space governed by the gauge action and controlled by the inverse bare gauge coupling, $\beta_L=2\,N/g_0^2$, %instead of the bare fermion mass, $m_0$, or the hopping parameter, $\kappa=1/\of{2a\,m_0+8}$, 
these boundary conditions are usually well behaved~\cite{Luscher:1992an,Luscher:1993gh,Sint:1995ch,Luscher:1996sc} and can even be further improved~\cite{Klassen:1997jf} to match bulk actions with reduced UV-cutoff effects~\cite{Weisz:1982zw,Iwasaki:1983iya,Luscher:1984xn,Sheikholeslami:1985ij}.

The boundary conditions can, however, cause problems in theories in which the fermionic action has a dominant effect on the UV physics, or in regions of parameter space of QCD-like theories, where the gauge action becomes sub-dominant. The latter can be the case, if the inverse bare gauge coupling, $\beta_L$, is too small, so that the Wilson gauge action from Eq.~\eqref{eq:wilsongaugeaction} does no longer sufficiently constrain the values of the traces of the plaquette variables in Eq.~\eqref{eq:plaquettevar}. Then, the behavior of the gauge field at the plaquette-level (i.e. UV-cutoff scale) is dictated by the interaction of the gauge field with the fermions.

\subsection{Why do the boundary conditions fail?}\label{ssec:failureofDBC}
It is well known that the Wilson and Staggered fermion actions induce a term which is proportional to the plaquette gauge action~\cite{Hasenfratz:1993az,Blum:1994xb,deForcrand:2012vh}, and therefore introduce a shift in the effective inverse gauge coupling: 
\[
\beta_{L,\mathrm{eff}}=\beta_{L}+\Delta\beta_L\ ,\label{eq:betashift}
\]
where $\Delta\beta_L\propto N_f$. If the bare gauge coupling is very strong, the shift $\Delta\beta_L$ can dominate $\beta_{L,\mathrm{eff}}$.

\begin{figure}[h]
\centering
\begin{tikzpicture}[scale=0.4,nodes={inner sep=0}]
  \pgfpointtransformed{\pgfpointxy{1}{1}};
  \pgfgetlastxy{\vx}{\vy}
  \begin{scope}[node distance=\vx and \vy]
    \foreach \i in {0,...,7} {
        \draw [thin,gray] (\i,0) -- (\i,8)  node[solid,black,above] at (\i,7.3) {};
    }
    \foreach \i in {0,...,8} {
        \draw [thin,gray] (0,\i) -- (8,\i) node[solid,black,left] at (0.7,\i) {};
    }
        
    \draw[very thick,black] (0,0) -- (8,0) node[black,right=0.1,pos=1,scale=1.1] {$x_4=0$};
    \draw[very thick,black] (0,8) -- (8,8) node[black,right=0.1,pos=1,scale=1.1] {$x_4=N_t$};

    \foreach \i in {0,...,7} {
        \draw [line width=1.5pt,dash pattern=on 2pt off 1.5pt,orange!95!black] (\i,8-0.05) -- (\i,8-0.95);
    }
    \foreach \i in {0,...,7} {
        \node[circle,fill=red!80!black,minimum size=5pt] at (\i,8) {};
    }
    \foreach \i in {0,...,7} {
        \draw[fill=blue!80!black] (\i+0.1,8-0.9) rectangle ++(0.8,0.8);
    }  
    \foreach \i in {0,...,7} {
        \draw [line width=1.5pt,dash pattern=on 2pt off 1.5pt,orange!95!black] (\i,0.05) -- (\i,0.95);
    }
    \foreach \i in {0,...,7} {
        \node[circle,fill=red!80!black,minimum size=5pt] at (\i,0) {};
    }
    \foreach \i in {0,...,7} {
        \draw[fill=blue!80!black] (\i+0.1,0.1) rectangle ++(0.8,0.8);
    }   
    
    \draw[->] (-1,0) -- (-1,2) node[black,thick,left=0.1,pos=0.5,scale=1.1] {$x_4$};
    \draw[->] (0,-1) -- (2,-1) node[black,thick,below=0.1,pos=0.5,scale=1.1] {$x_i$};
  \end{scope}
\end{tikzpicture}
\caption{Illustration of the problem caused by the trivial temporal Dirichlet boundary conditions (TDBs) from Fig.~\ref{fig:normalhomdiribc}: as the fermion fields are required to vanish on the boundaries at $x_4=0,N_t$, fermion hopping is prevented along the temporal links that connect to the boundaries (orange, dashed links). Consequently, the effect of the fermion fields on gauge links that are part of temporal boundary plaquettes (blue squares) is significantly smaller on gauge links deeper in the bulk.}
\label{fig:normalhomdiribcfail}
\end{figure}
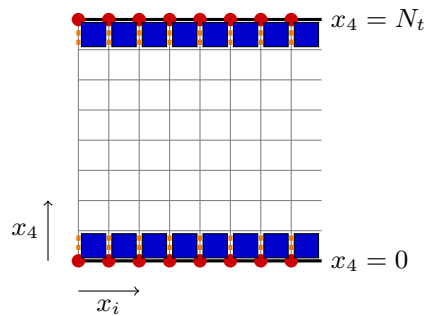

The reason why the TDBs from Fig.~\ref{fig:normalhomdiribc} may cause problems is illustrated in Fig.~\ref{fig:normalhomdiribcfail}. The requirement for the fermion fields to vanish on the $x_4=0$ and $x_4=N_t$ boundaries implies that the fermion ``hopping terms'' along temporal links connecting to either of the boundaries (dashed, orange lines) vanish.  This means that there are no fermion loops around temporal plaquettes which touch the boundary (blue squares) and hence the increase in effective $\beta_L$ due to fermions is strongly suppressed.

Thus, the boundary plaquettes are effectively at much stronger coupling than the plaquettes in the bulk. As will be demonstrated below, this has an effect in the measurement of the gradient flow coupling.
If the bare lattice coupling is weak, the shift can be compensated in perturbation theory \cite{Sint:1995ch,Luscher:1996sc,Luscher:1996vw,Karavirta:2011mv,Karavirta:2012qd,Hietanen:2014lha}. At strong coupling, however, the perturbative improvement is not sufficient.

\subsection{Example: severe boundary effects in $\SU{2}$ with $N_f=24$ massless fermions}\label{ssec:exampleoffailing}
To provide a concrete example, we consider here a $\SU{2}$ lattice gauge theory with Wilson plaquette gauge action from Eq.~\eqref{eq:wilsongaugeaction} and $N_f=24$ massless (in terms of PCAC mass~\cite{Luscher:1996vw}), fundamental Wilson-clover fermions with action as in Eq.~\eqref{eq:wilsonfermionaction} and with Sheikholeslami-Wohlert coefficient $c_{\mathrm{SW}}=1$~\cite{Rantaharju:2015yva}. The fermions couple to hypercubically truncated stout-smeared (HEX-smeared)~\cite{Capitani:2006ni} gauge links. 
The theory is not asymptotically free, but it is still of interest to observe the behaviour of the theory at strong coupling in the light of asymptotic safety~\cite{Antipin:2017ebo,Dondi:2019ivp,Leino:2019qwk}.

In previous lattice studies of the model \cite{Leino:2019qwk,Rantaharju:2021iro,Rindlisbacher:2021hhh} the requirement of strong effective coupling (small $\beta_{L,\mathrm{eff}}$) forced one to use even negative values for the bare inverse gauge coupling $\beta_L$. Indeed, in ref.~\cite{Leino:2019qwk} it was estimated that $\beta_{L,\mathrm{eff}} \approx 0.7$ with the bare $\beta_L = -0.3$. 
  
\begin{figure}[h]
    \centering
    \let\fwidth\linewidth
    \begin{tikzpicture}[scale=1.,nodes={inner sep=0},every node/.style={transform shape}]
      \node [above right] at (0,0) {\includegraphics[width=0.485\fwidth,keepaspectratio]{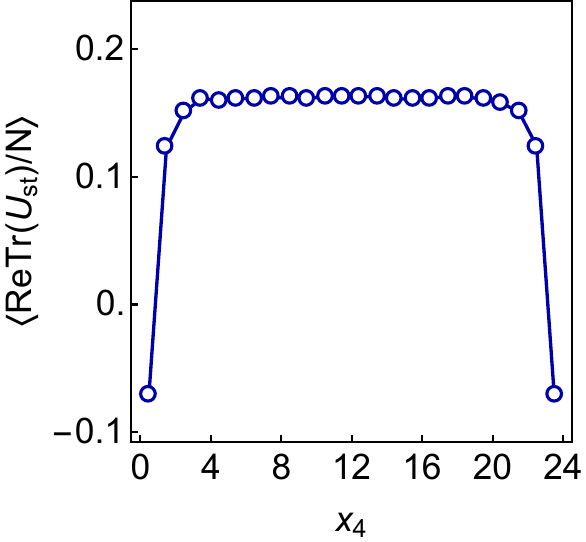}};
      \node [above right] at (0.5\fwidth,0) {\includegraphics[width=0.485\fwidth,keepaspectratio]{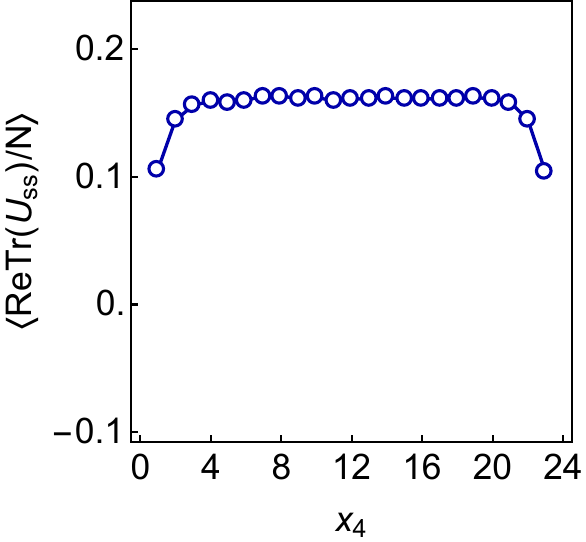}};
      \node [above right] at(0,3.35) {a)};
      \node [above right] at(0.5\fwidth,3.35) {b)};
    \end{tikzpicture}
    \caption{Average temporal (a) and spatial (b) plaquette per time slice (labeled by $x_4$) for a $\SU{2}$ lattice gauge theory with $N_f=24$ massless (w.r.t. PCAC mass) Wilson-clover flavors, measured at $\beta_L=-0.3$ on a $L^4=24^4$ lattice with the boundary conditions from Fig.~\ref{fig:normalhomdiribc}. The spatial plaquette values at $x_4=0,24$ are not shown as the boundary conditions require them to be unity. The temporal plaquettes at the boundary have a negative value, as expected for negative bare $\beta_L$ if the fermion-induced $\Delta\beta_L$ (cf. Eq.~\eqref{eq:betashift}) is absent.}
    \label{fig:boundaryplaqeffect}
\end{figure}

The boundary effect can be clearly observed in Fig.~\ref{fig:boundaryplaqeffect}, which shows measurements of the average temporal (panel (a)) and average spatial (panel (b)) plaquettes per time-slice, obtained from a HMC simulation of the $\SU{2}$, $N_f=24$ theory at $\beta_L=-0.3$. The temporal plaquettes at the boundary have a negative value, as expected for negative $\beta_L$ in absence of the fermion-induced shift $\Delta\beta_L$ from Eq.~\eqref{eq:betashift}.

While Fig.~\ref{fig:boundaryplaqeffect} seems to indicate that
the boundary effects vanish quickly with increasing distance from the boundary, their presence can in fact severely affect measurements of bulk quantities. The reason for this is, that the strongly fluctuating gauge field at the boundary can easily induce topology changes via instanton/anti-instanton creation/annihilation. Although these instantons mostly remain localized near the boundary during the simulation, the Dirichlet boundary conditions do not allow the instantons to disappear through the boundary during the gradient flow. Consequently, the excess energy stored within these instantons has to dissipate in the bulk under the action of the gradient flow, either in the form of remnant heat instanton/anti-instanton pairs that annihilate under the action of the flow, or in the form of stable instantons, which the flow pushes away from the flat TDBs.

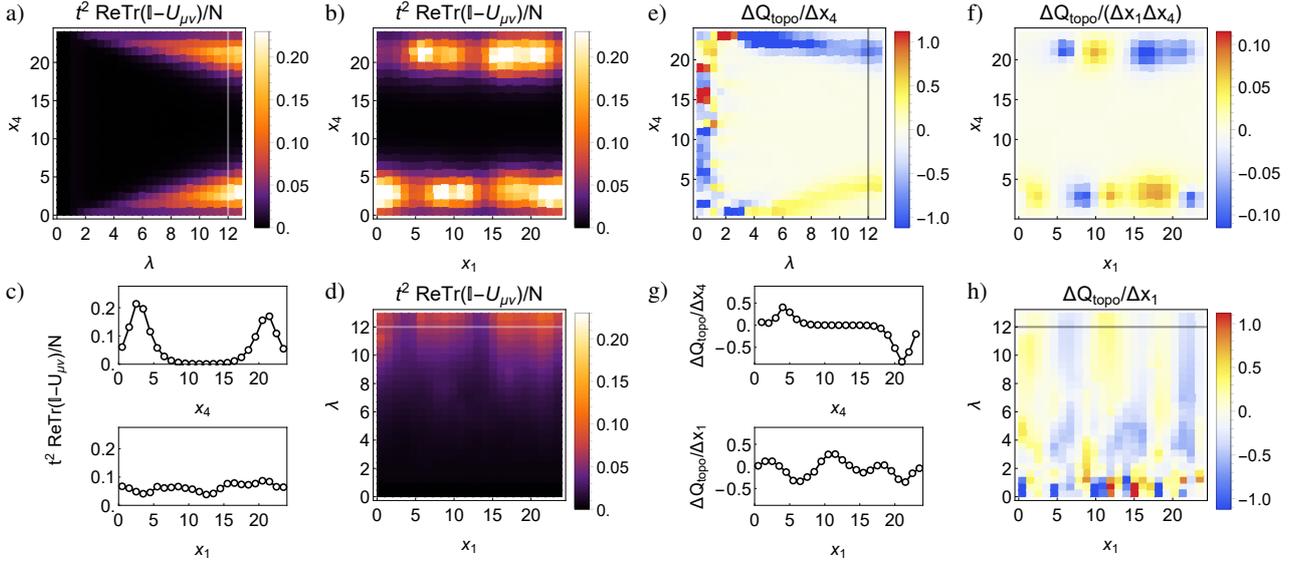
\begin{figure*}[!htb]
    \centering
    \let\fwidth\linewidth
    \begin{tikzpicture}[scale=1.,nodes={inner sep=0},every node/.style={transform shape}]
      \node [above right] at (0,0) {\includegraphics[width=\fwidth,keepaspectratio]{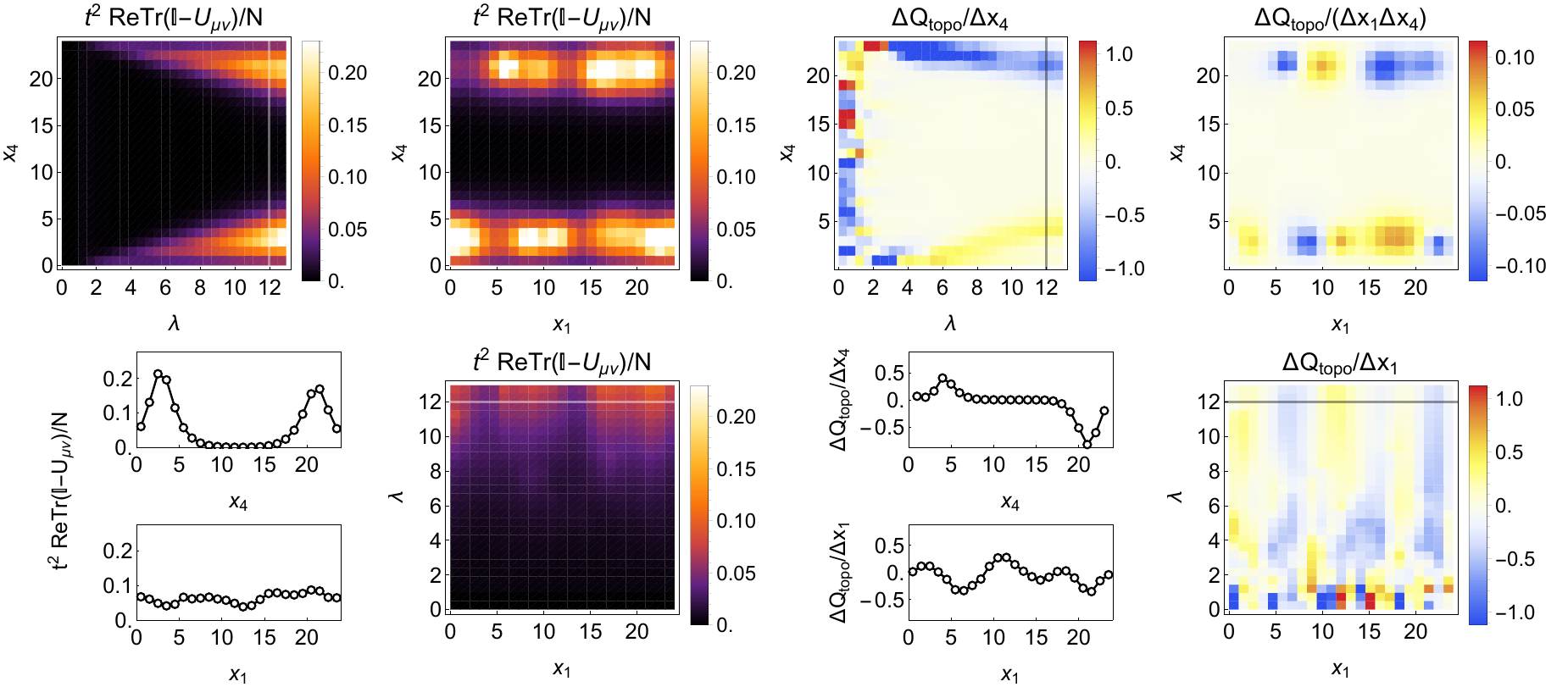}};
      %\draw[draw=red!50,thick,fill=red!10,opacity=0.6] (3.22,0.9) rectangle ++(2.45,2.7);
      \node [above right] at(0,7.1) {a)};
      \node [above right] at(4.2,7.1) {b)};
      \node [above right] at(0,3.4) {c)};
      \node [above right] at(4.2,3.4) {d)};
      
      \node [above right] at(8.45,7.1) {e)};
      \node [above right] at(12.65,7.1) {f)};
      \node [above right] at(8.45,3.4) {g)};
      \node [above right] at(12.65,3.4) {h)};
    \end{tikzpicture}
    \caption{Illustration of how the rescaled plaquette action density (panels (a)-(d)) and topological charge distribution (panels (e)-(h)) in a typical configuration, obtained at $\beta_L=-0.3$ with the homogeneous temporal Dirichlet boundaries (TDBs) from Fig.~\ref{fig:normalhomdiribc}, evolve under the gradient flow. The Dirichlet boundaries are located at $x_4=0,N_t$, where $N_t=24$, and $\lambda=\sqrt{8\,t}$ is the gradient flow scale after flow time $t$. Panels (a) and (d) show, respectively, how the average re-scaled plaquette action per $x_4$- and $x_1$-slice changes as function of $\lambda$, while panel (b) shows how the quantity is distributed as function of both, $x_1$ and $x_4$ at the value of $\lambda$ that is indicted by the grey lines in panels (a) and (d). The plots in (c) represent the cross-sections along the grey lines in (a) and (d). The panels (e)-(h) represent analogously the data for the corresponding topological charge.}
    \label{fig:boundaryinstantons}
\end{figure*}

This is visualized in Fig.~\ref{fig:boundaryinstantons}, which shows how the rescaled plaquette action density from Eq.~\eqref{eq:plqsdens} (panels (a)-(d)) and the topological charge distribution from Eq.~\eqref{eq:topodens} (panels (e)-(h)) in a typical configuration at $\beta_L=-0.3$ evolve under the gradient flow. 

Panels (a) and (d) show, respectively, how the average rescaled plaquette action per $x_4$- and $x_1$-slice changes as function of flow scale $\lambda=\sqrt{8\,t}$, with $t$ being the gradient flow time, while panel (b) shows how the quantity is distributed as function of both, $x_1$ and $x_4$ at the value of $\lambda$ that is indicated by the grey lines in panels (a) and (d) (i.e. at $\lambda=N_t/2=12$). The plots in (c) represent, respectively, the cross-sections along the grey lines in (a) and (d). 

\begin{figure}[h]
    \centering
    \includegraphics[width=0.8\linewidth,keepaspectratio]{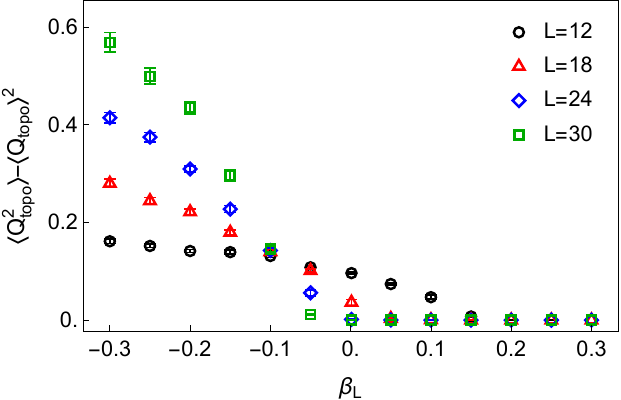}
    \caption{Topological susceptibility form Eq.~\eqref{eq:toposusc} for $\SU{2}$ lattice gauge theory with $N_f=24$ massless Wilson-clover fermions as function of $\beta_L$ for four different system sizes $V=L^4$ with $L=12,18,24,30$ and homogeneous temporal Dirichlet boundary conditions (TDBs). The non-zero topological susceptibility at strong coupling is merely due to the artificial instanton/anti-instanton production at the boundaries.}
    \label{fig:toposuscartifacts}
\end{figure}

The panels (e)-(h) represent analogously the data for the corresponding topological charge per $x_4$- and $x_1$- slice as function of $\lambda$. These show that with increasing flow scale the instantons indeed appear near the boundaries and then move slowly away from them. No instantons show up directly from within the bulk. The latter suggests that in Fig.~\ref{fig:toposuscartifacts} the non-zero topological susceptibility at small $\beta_L$ is primarily a boundary artifact. Since $\SU{2}$ gauge theory with $N_f=24$ fundamental, massless fermion flavors is an IR-free theory~\cite{Rantaharju:2021iro,Rindlisbacher:2021hhh}, one would indeed expect the topological susceptibility to vanish.

The presence of the artificial instantons slows down the decay of the action density under the effect of the gradient flow, so that $t^2\Re\Tr\of{\id-U_{\mu\nu}}$ in panels (a)-(d) of Fig.~\ref{fig:boundaryinstantons} grows rapidly as function of flow time $t$ in the region where the instantons are located. 

\begin{figure}[h]
\centering
\includegraphics[width=0.8\linewidth,keepaspectratio]{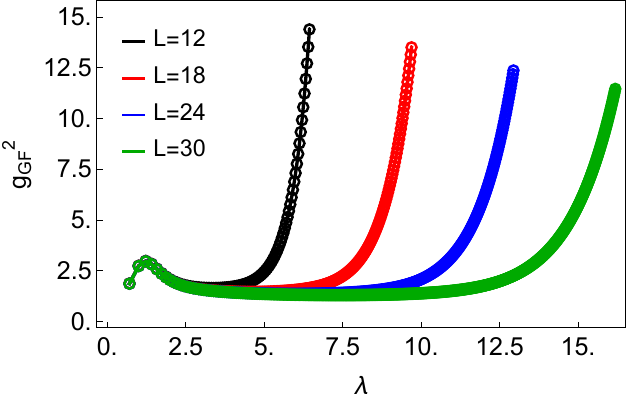}
\caption{Gradient flow coupling, $g_{\mathrm{GF}}^2$, from Eq.~\eqref{eq:gradientflow}, as function of flow scale, $\lambda$, for $\SU{2}$ lattice gauge theory with $N_f=24$ massless Wilson-clover fermions at $\beta_L=-0.3$ for four different system sizes $V=L^4$ with $L=12,18,24,30$ and homogeneous temporal Dirichlet boundary conditions (TDBs). The rapid increase of $g_{\mathrm{GF}}^2$ when $\lambda$ approaches $L/2$ is a pure finite-volume artifact, caused by the naive implementation of the TDBs.}
\label{fig:gffinitevolartifacts}
\end{figure}

Although the upper plot of panel (c) in Fig.~\ref{fig:boundaryinstantons} indicates that the peaks in the action-distribution are still around $x_4=3$ and $x_4=21$ for $\lambda=12$, the excess action due to instantons has indeed already reached the central time slice and is the reason for severe IR-cutoff effects in the gradient flow running coupling shown in Fig.~\ref{fig:gffinitevolartifacts}: as the gradient flow scale, $\lambda$, approaches the the value $N_t/2$, the gradient flow coupling, $g_{\mathrm{GF}}^2$, from Eq.~\eqref{eq:gradientflow} starts to increase rapidly. The theory is infrared free, and the physical coupling is expected to decrease with the distance scale ($\lambda$).
The rapid increase of the coupling is an unphysical finite volume effect, which can be seen from the fact that the length scale where the coupling starts to increase grows as the volume increases. This effect cannot be compensated by the finite volume correction discussed in Eq.~\eqref{eq:gradientflow}.

\section{Improved homogeneous Dirichlet boundary conditions}\label{sec:imprdirichletbc}
In the previous section we have seen that the homogeneous TDBs from Fig.~\ref{fig:normalhomdiribc} can cause sever problems in lattice simulations of gauge-fermion theories  if the simulation parameters are such that the physics at the UV-cutoff scale is predominantly governed by the fermionic instead of the gauge action. These are caused by the fact that the boundary conditions strongly reduce the effect of the fermion action on the plaquettes touching the boundary.

\begin{figure}[h]
\centering
\begin{tikzpicture}[scale=0.4,nodes={inner sep=0}]
  \pgfpointtransformed{\pgfpointxy{1}{1}};
  \pgfgetlastxy{\vx}{\vy}
  \begin{scope}[node distance=\vx and \vy]
    \foreach \i in {0,...,7} {
        \draw [thin,gray] (\i,0) -- (\i,8)  node[solid,black,above] at (\i,7.3) {};
    }
    \foreach \i in {0,...,8} {
        \draw [thin,gray] (0,\i) -- (8,\i) node[solid,black,left] at (0.7,\i) {};
    }
        
    \draw[very thick,black] (0,0) -- (8,0) node[black,right=0.1,pos=1,scale=1.1] {$x_4=0$};
    \draw[very thick,black] (0,8) -- (8,8) node[black,right=0.1,pos=1,scale=1.1] {$x_4=N_t$};

    \foreach \i in {0,...,7} {
        \draw [very thick,blue!80!black] (\i+0.05,8) -- (\i+0.95,8);
        \draw [very thick,blue!80!black] (\i+0.05,7) -- (\i+0.95,7);
        \draw [very thick,blue!80!black] (\i,7.05) -- (\i,7.95);
    }
    \foreach \i in {0,...,7} {
        \node[circle,fill=red!80!black,minimum size=5pt] at (\i,8) {};
    }    
    \foreach \i in {0,...,7} {
        \draw [very thick,blue!80!black] (\i+0.05,0) -- (\i+0.95,0);
        \draw [very thick,blue!80!black] (\i+0.05,1) -- (\i+0.95,1);
        \draw [very thick,blue!80!black] (\i,0.05) -- (\i,0.95);
    }
    \foreach \i in {0,...,7} {
        \node[circle,fill=red!80!black,minimum size=5pt] at (\i,0) {};
    }    
    
    \draw[->] (-1,0) -- (-1,2) node[black,thick,left=0.1,pos=0.5,scale=1.1] {$x_4$};
    \draw[->] (0,-1) -- (2,-1) node[black,thick,below=0.1,pos=0.5,scale=1.1] {$x_i$};
  \end{scope}
\end{tikzpicture}
\caption{Improved, ``thick'' homogeneous temporal Dirichlet boundary conditions:    all link variables that are part of temporal boundary plaquettes are set to the identity matrix (blue links). The fermion fields are required to vanish on the boundary (red dots).}
\label{fig:thickhomdiribc}
\end{figure}
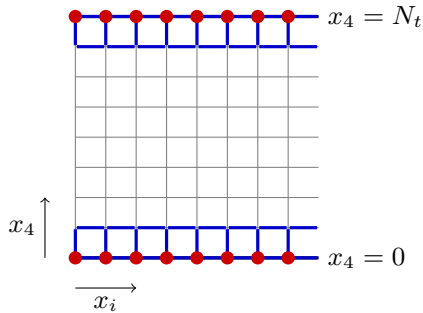

A simple way to regulate the behaviour of the gauge field on the temporal boundary plaquettes 
%form behaving randomly and affecting negatively the physics in the bulk of the system, 
is to simply freeze the involved gauge links, by setting them all to unity:
\begin{subequations}
\begin{align}
U_{x,i=1,2,3}&=\id\quad\text{if}\quad x_4\in\cof{0,1,N_t-1,N_t}\ , \label{eq:improvedA}\\
U_{x,4}&=\id\quad\text{if}\quad x_4\in\cof{0,N_t-1}\ .
\label{eq:improvedB}
\end{align}
The fermions are still required to vanish on the boundaries:
\begin{align}
\psi_{x}&=0\quad\text{if}\quad x_4\in\cof{0,N_t}\ ,\\
\bar{\psi}_{x}&=0\quad\text{if}\quad x_4\in\cof{0,N_t}\ .
\end{align}
\end{subequations}
These boundary conditions, to which we will refer as ``thick'' improved homogeneous TDBs, are graphically depicted in Fig.~\ref{fig:thickhomdiribc}. 
They are particularly easy to implement in a parallel HMC code, as one can simply set the force terms for the blue-marked links in Fig.~\ref{fig:thickhomdiribc} to zero during the gauge update. The side effect of implementing the thick TDBs in this way is, that the active gauge field lattice volume gets reduced from $N_s^3\times N_t$ to $V_A=N_s^3\times \of{N_t-2}$, which can be compensated by increasing the original $N_t$ by two units.

Physically equivalent boundary condition to Eqs.~(\ref{eq:improvedA}--\ref{eq:improvedB}) can also be obtained by \kr{using standard ``thin'' TDBs and} setting $\beta_L$ to infinity for plaquettes which include link variables on $x_4=0$ or $x_4=N_t$. This forces the blue links in Fig.~\ref{fig:thickhomdiribc} to become gauge equivalent with the unit matrix. We note that the perturbative 1-loop boundary improvement due to fermions also increases the effective inverse coupling for the boundary plaquette~\cite{Sint:1995ch,Luscher:1996sc,Luscher:1996vw}, but at strong coupling this perturbative correction is insufficient.
The goal of the perturbative improvement is also slightly different from the ``thick'' TDBs introduced here: the perturbative improvement aims to compensate the boundary lattice effects while keeping the original active volume from $x_4=0$ to $N_t$.  In contrast, the ``thick'' boundary conditions Eqs.~(\ref{eq:improvedA}--\ref{eq:improvedB}) exclude the boundary layers from the physical volume, reducing the active volume to the interval $1\le x_4 \le N_t-1$.
   
To demonstrate that the improved, ``thick'' TDBs, just introduced in Sec.~\ref{sec:imprdirichletbc}, prevent the formation of the boundary artifacts discussed in Sec.~\ref{sec:dirichletbcsc}, we repeat some of the $\SU{2}$, $N_f=24$ simulations and the related analysis.

\begin{figure}[h]
    \centering
    \let\fwidth\linewidth
    \begin{tikzpicture}[scale=1.,nodes={inner sep=0},every node/.style={transform shape}]
      \node [above right] at (0,0) {\includegraphics[width=0.485\fwidth,keepaspectratio]{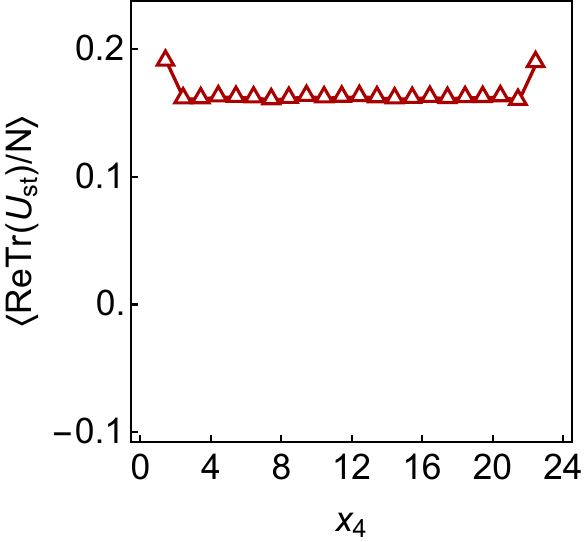}};
      \node [above right] at (0.5\fwidth,0) {\includegraphics[width=0.485\fwidth,keepaspectratio]{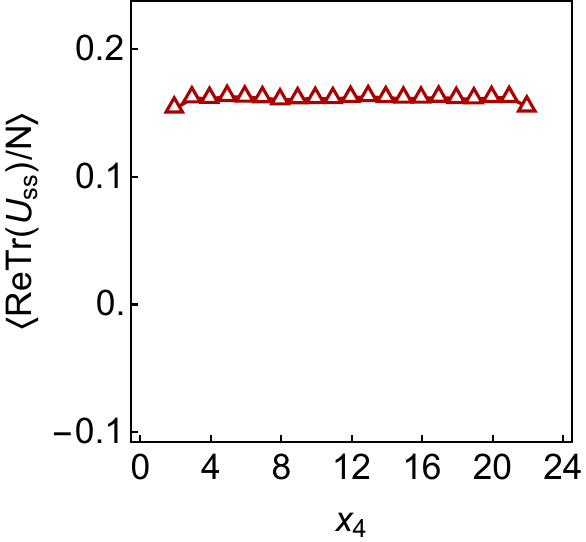}};
      \node [above right] at(0,3.35) {a)};
      \node [above right] at(0.5\fwidth,3.35) {b)};
    \end{tikzpicture}
    \caption{Average temporal (a) and spatial (b) plaquette per time slice (labeled by $x_4$) for a $\SU{2}$ lattice gauge theory with $N_f=24$ massless Wilson-clover flavors, measured at $\beta_L=-0.3$ on a $L^4=24^4$ lattice with the improved temporal Dirichlet boundary conditions from Fig.~\ref{fig:thickhomdiribc}. The unity plaquette values from within the boundary layers are not plotted.}
    \label{fig:boundaryplaqimpr}
\end{figure}

\begin{figure*}[htb]
    \centering
    \let\fwidth\linewidth
    \begin{tikzpicture}[scale=1.,nodes={inner sep=0},every node/.style={transform shape}]
          \node [above right] at (0,0) {\includegraphics[width=\fwidth,keepaspectratio]{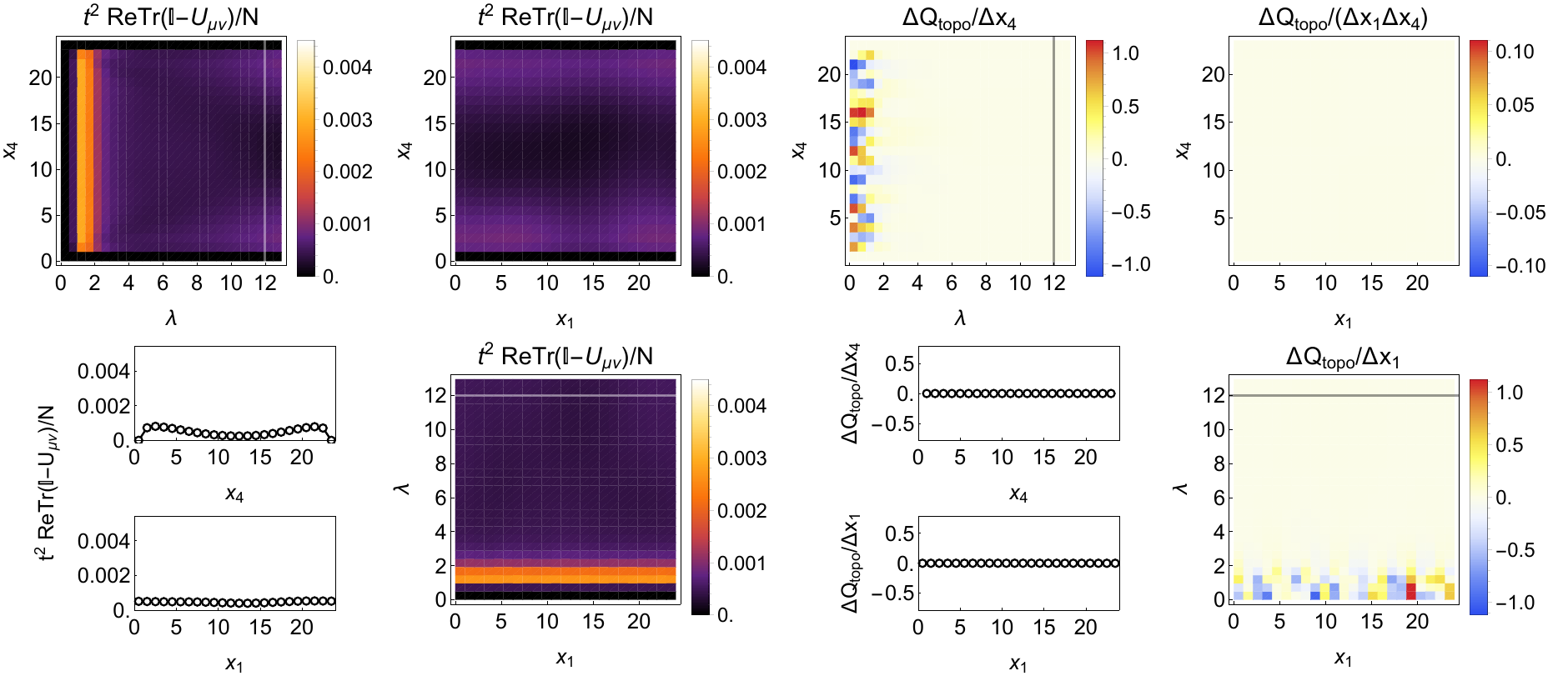}};
          %\draw[draw=red!50,thick,fill=red!10,opacity=0.6] (3.22,0.9) rectangle ++(2.45,2.7);
          \node [above right] at(0,7.) {a)};
          \node [above right] at(4.2,7.) {b)};
          \node [above right] at(0,3.32) {c)};
          \node [above right] at(4.2,3.32) {d)};
          
          \node [above right] at(8.45,7.) {e)};
          \node [above right] at(12.65,7.) {f)};
          \node [above right] at(8.45,3.32) {g)};
          \node [above right] at(12.65,3.32) {h)};
    \end{tikzpicture}
    \caption{Illustration of how the rescaled plaquette action density (panels (a)-(d)) and topological charge distribution (panels (e)-(h)) in a typical configuration, obtained at $\beta_L=-0.3$ with thick homogeneous temporal Dirichlet boundaries, evolve under the gradient flow. The data is represented analogously as in Fig.~\ref{fig:boundaryinstantons}.}
\label{fig:boundaryinstantonabsence}
\end{figure*}
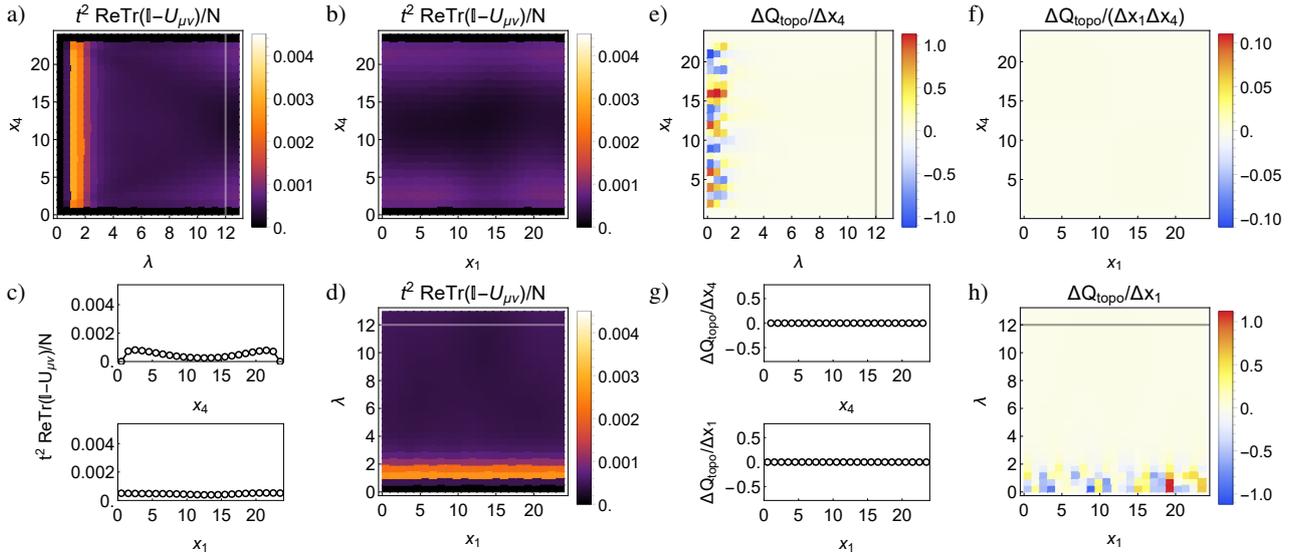

With the ``thick'' temporal boundary conditions from Fig.~\ref{fig:thickhomdiribc}, the temporal boundary plaquettes, that do not experience the fermion field, are set to unity, so that they cannot act up and give rise to any sort of non-trivial artifacts that could alter the physics in the bulk. Comparing Fig.~\ref{fig:boundaryplaqimpr} with Fig.~\ref{fig:boundaryplaqeffect}, we see that the ``thick'' TDBs affect the average plaquette values per time slice much less than the normal TDBs did. In particular, the plaquettes near the boundary are no longer significantly shifted to smaller values, indicating that the boundary layers are no longer at stronger effective coupling than the bulk. As a consequence, the excess instanton production we observed with the ordinary TDBs (cf. in Fig.~\eqref{fig:boundaryinstantons}), should now be absent. 
Fig.~\ref{fig:boundaryinstantonabsence} show that the latter is indeed the case: panels (e)-(h) show that no excess instantons show up near the temporal boundaries and the rescaled action density in panels (a)-(d) does no longer blow up near the boundary as in Fig.~\eqref{fig:boundaryinstantons}.

\begin{figure}[h]
    \centering
    \includegraphics[width=0.8\linewidth,keepaspectratio]{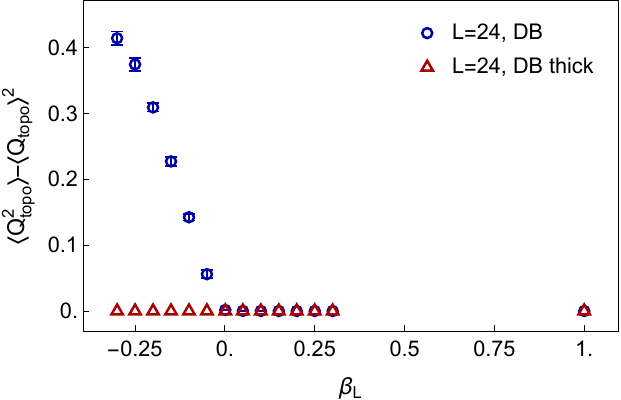}
    \caption{Topological susceptibility from Eq.~\eqref{eq:toposusc} for $\SU{2}$ lattice gauge theory with $N_f=24$ massless Wilson-clover fermions as function of $\beta_L$ on a $L^4$ lattice with $L=24$. The blue circles show the results obtained with the normal TDBs from Fig.~\ref{fig:normalhomdiribc}, and the red triangles the corresponding results for the improved, ``thick'' boundary conditions (DB thick) from Fig.~\ref{fig:thickhomdiribc}. The non-zero topological susceptibility at strong coupling for DB is merely due to the artificial instanton/anti-instanton production at the boundaries.}
    \label{fig:toposusccomp}
\end{figure}

In Fig.~\ref{fig:toposusccomp} we compare data for the topological susceptibility from Eq.~\eqref{eq:toposusc} on a $L=24$ lattice, obtained with normal and ``thick'' TDBs. As can be seen, with the ``thick'' TDBs, the topological susceptibility remains zero all the way down to $\beta_L=-0.3$, indicating that the non-zero topological susceptibility, obtained with the normal TDBs, is a consequence of the artificial instanton/anti-instanton production at the boundary.

\begin{figure}[h]
\centering
\let\fwidth\linewidth
{\small $\beta_L=-0.3$}\\[0pt]
\includegraphics[width=0.8\fwidth,keepaspectratio]{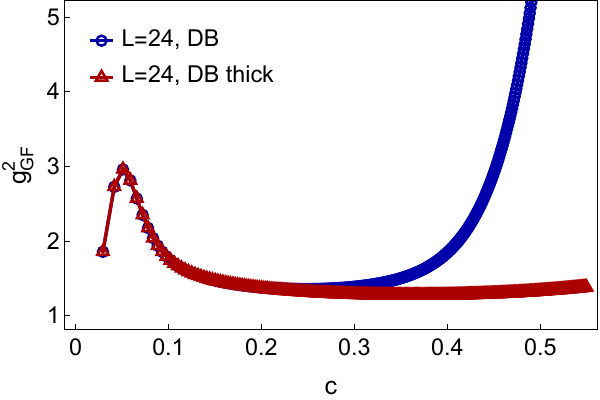}\\[5pt]
{\small $\beta_L=1.0$}\\[0pt]
\includegraphics[width=0.8\fwidth,keepaspectratio]{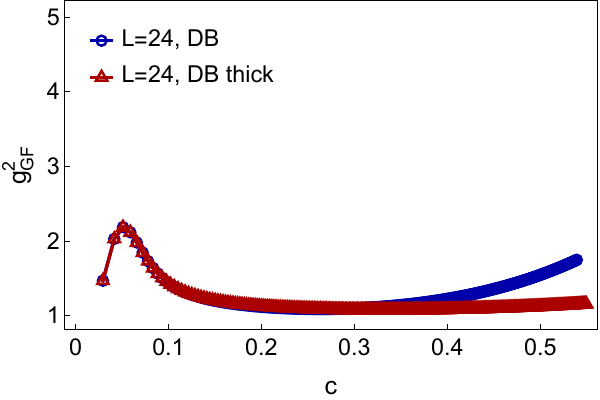}
\caption{Gradient flow (GF) running coupling from Eq.~\eqref{eq:gradientflow} for $\SU{2}$ lattice gauge theory with $N_f=24$ massless Wilson-clover fermions at $\beta_L=-0.3$ (top) and $\beta_L=1.0$ (bottom) on a $L^4$ lattice with $L=24$ as function of $c=\lambda/L$, where $\lambda=\sqrt{8\,t}$ is the flow scale at flow time $t$. The blue data corresponds to the GF running coupling for a system with normal temporal Dirichlet boundary conditions (cf. Fig.~\ref{fig:normalhomdiribc}), while the red data shows the GF running coupling for a corresponding system with improved, ``thick'' boundary conditions (cf. Fig.~\ref{fig:thickhomdiribc}).}
\label{fig:runningcouplingcomp}
\end{figure}

In Fig.~\ref{fig:runningcouplingcomp} the corresponding gradient flow running coupling, $g_{\mathrm{GF}}^2$, from Eq.~\eqref{eq:gradientflow} is shown as function of $c=\lambda/L$, for normal (blue data) and improved ``thick'' (red data) TDBs for $\beta_L=-0.3$ (top) and $\beta_L=1.0$ (bottom). For the improved ``thick'' TDBs, the finite volume correction in Eq.~\eqref{eq:gradientflow} has been adjusted to incorporate the fact that the active volume for the gauge field is reduced by two time slices $N_t\to N_t-2\neq L$. As discussed in Sec.~\ref{ssec:exampleoffailing}, in the system with normal boundary conditions, the instantons and anti-instantons produced at the boundary for $\beta_L=-0.3$ slow down the decay of the action density in the gradient flow-evolved gauge field, which spoils the definition of the running GF coupling. As soon as the flow scale, $\lambda$, of the gradient flow is sufficiently large, so that the flowed field from near the boundary reaches the central time slice at $x_4=L/2$, where the clover action for the gradient flow coupling is measured, the latter starts to increase rapidly. With the ``thick'' boundary conditions, there are no boundary instantons that interfere with the gradient flow, and the definition of the gradient flow coupling remains valid when the relative flow scale $c=\lambda/L$ approaches the value $c=0.5$. The gradient flow coupling therefore remains flat, as expected for an IR-free theory.
For $\beta_L=1.0$, the gauge action is by itself, i.e. also in the absence of fermions, able to sufficiently constrain the temporal boundary plaquettes in the system with naive DB boundary conditions, so that no excess instantons get produced. The gradient flow coupling with normal boundary conditions behaves therefore almost like the one with the ``thick'' boundary conditions. The remnant additional increase in the gradient flow coupling with naive DB boundary conditions can be blamed to the fact that the normal boundary conditions nevertheless tend to increase the action in the near-boundary region, while the ``thick'' boundary conditions tend to decrease it (compare Fig.~\ref{fig:boundaryplaqeffect} and Fig.~\ref{fig:boundaryplaqimpr}).

\section{Conclusions and outlook}\label{sec:conclusion}
We have introduced modified homogeneous temporal Dirichlet boundary conditions (TDBs) for lattice studies of gauge-fermion theories at strong bare lattice gauge coupling. The boundary conditions are designed to reduce lattice artifacts caused by the TDBs, in particular for observables measured using gradient flow. We demonstrated that the new boundary conditions dramatically reduce the finite volume (IR-cutoff) artifacts in lattice simulations of $\SU{2}$ gauge theory with $N_f=24$ massless Wilson-clover fermions. Similar improvements are expected in other theories where the bare lattice gauge coupling is large. 

Large values of the bare lattice gauge coupling have to be faced e.g. in studies of asymptotically free theories which have an infrared conformal fixed point (or are close to having one). In order to have strong effective coupling at the scale of the lattice size, it is then necessary to use large bare lattice gauge coupling, due to the very slow running of the coupling with increasing length scale. 
Large bare coupling values can lead to harmful discretization artifacts in the lattice gauge field. There have been recent proposals to reduce these artifacts by modifying the lattice gauge action~\cite{Hasenfratz:2011xn,Hasenfratz:2021zsl,Rindlisbacher:2023qjg} and to reach stronger effective coupling values at weak bare coupling by introducing heavy auxiliary scalar fields~\cite{Hasenfratz:2021zsl}. 
The interplay of these proposals and the improved Dirichlet boundary conditions will be the subject of a future study.

\section{Acknowledgement}\label{sec:aknwldg}
The support of the Academy of Finland grants 345070 and 320123 is acknowledged. T.~R. is supported by the Swiss National Science Foundation (SNSF) through the grant no.~TMPFP2\_210064. The authors wish to acknowledge CSC - IT Center for Science, Finland, for computational resources.

\end{document}